\documentclass[aps,prl,twocolumn,floats,showpacs]{revtex4}
\usepackage{epsfig}
\usepackage{amsmath}
\usepackage{amssymb}
\begin{document}
\title{Magnetically tunable Kondo -- Aharonov-Bohm  effect in triangular quantum
dot}
\author{T. Kuzmenko$^1$, K. Kikoin$^1$ and Y. Avishai$^{1,2}$}
\affiliation{$^1$Department of Physics, $^2$ Ilse Katz Center for
Nano-Technology, Ben-Gurion University of the Negev, Beer-Sheva
84105, Israel}
\begin{abstract}
The role of discrete orbital symmetry in nanoscopic physics is
manifested in a system consisting of three identical quantum dots
forming an equilateral triangle. Under a perpendicular magnetic
field, this system demonstrates a unique combination of Kondo and
Aharonov-Bohm features due to an interplay between continuous
(spin-rotation $SU(2)$) and discrete (permutation $C_{3v}$)
symmetries, as well as $U(1)$ gauge invariance. The conductance as
a function of magnetic flux displays sharp enhancement or complete
suppression depending on contact setups.
\end{abstract}
 \maketitle

Experimental analysis of the Kondo effect in simple quantum dots
(QD) \cite{KKK} treats the electron as a local spin $1/2$ magnetic
moment devoid of orbital degrees of freedom. These are absent also
in theoretical discussions of the Kondo effect in composite
structures consisting of two or three dots
\cite{DD,KAv,Hofsh,KuKA}. However, orbital effects, which play a
crucial role in real metals \cite{CqS, NB}, become relevant also
in mesoscopic physics, e.g., when a QD is fabricated in a {\em
ring} geometry, having discrete point symmetries. At low
temperature it can serve both as a Kondo-scatterer and as a
peculiar Aharonov-Bohm (AB) interferometer, since the magnetic
flux affects the nature of the QD ground and excited states. The
simplest such system (three dots forming a triangle) has been
realized experimentally \cite{stop2, stop1}. Triangular trimer of
Cr ions on a gold surface was also studied \cite{Jam}. The orbital
symmetry of triangle is discrete.  It results in additional
degeneracies of the spectrum of trimer, which may be the source of
non Fermi liquid (NFL) regime \cite{Affleck}.

In the present work we analyze the physics of tunneling through a
triangular triple quantum dot (TTQD) in a magnetic field with one
electron shared by its three identical constituents (see Fig. 1).
It exhibits an interplay between continuous $SU(2)$ electron {\it
spin symmetry}, discrete {\it point symmetry} $C_{3v}$ and $U(1)$
{\it gauge invariance} of electron wave functions in an external
magnetic field. Its conductance is characterized by an unusual
dependence on the magnetic flux $\Phi$ through the triangle,
displayed by sharp peaks or narrow dips, depending on contact
geometry. In a 3-terminal geometry (Fig. 1a) the sharp peaks arise
since the magnetic field induces a symmetry crossover $SU(2)\to
SU(4)$. In a 2-terminal geometry (Fig. 1b) the Kondo tunneling is
modulated by
 AB interference, which blocks the source-drain cotunneling amplitude at certain
 flux values.
This Kondo-AB interplay should not be confused with that in
mesoscopic structures with QD as an element in the AB loop
\cite{AB}.\\
 A symmetric TTQD in contact with metallic leads is
described by the Hamiltonian $ H=H_{d}+H_{lead}+H_t,$ expressed in
terms of dot and lead operators $d_{j \sigma}, c_{j \sigma},$ with
$ j=1,2,3,$ and $ \sigma=\uparrow,\downarrow$.  $H_{d}$ describes
an isolated TTQD,
\begin{eqnarray}
&&H_{d}=\epsilon\sum_{j=1}^3\sum_{\sigma}d^\dagger_{j
\sigma}d_{j\sigma }+Q\sum_{j}n_{j\uparrow}n_{j\downarrow}\label{H-dot} \\
&& +Q'\sum_{\langle jl\rangle}\sum_\sigma n_{j\sigma}n_{l\sigma'}
+W\sum_{\langle jl\rangle}\sum_{\sigma}(d^\dagger_{j
\sigma}d_{l\sigma }+H.c.). \nonumber
\end{eqnarray}
Here $\langle jl \rangle=
\langle12\rangle,\langle23\rangle,\langle31\rangle$, $Q$ and $Q'$
are intradot and interdot
 charging energies ($Q\gg Q'$), and $W$ is the interdot tunneling
 amplitude. $H_{lead}$ describes electrons in the respective electrodes,
\begin{eqnarray}
H_{lead}&=&\sum_{j{\bf k}\sigma}\epsilon_{jk}c^\dagger_{j{\bf k}
\sigma}c_{j{\bf k}\sigma}, \label{H-l}
\end{eqnarray}
and $H_t$ is the tunneling Hamiltonian
\begin{eqnarray}
H_t=V\sum_{j{\bf k}\sigma}( c^\dagger_{j{\bf k}\sigma}
d_{j\sigma}+ H.c.).\label{H-tun}
\end{eqnarray}
\begin{figure}[htb]
\centering
\includegraphics[width=50mm,height=40mm,angle=0]{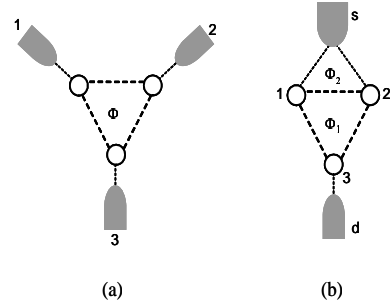}
\caption{Triangular triple quantum dot (TTQD) in three-terminal
(a) and two-terminal (b) configurations.}\label{figTQD}
\end{figure}
The dot energy $\epsilon$ is tuned by gate voltage in such a way
that the ground-state occupation of the isolated TTQD is ${\cal
N}=1$. Consider first a TTQD with three leads  and three identical
channels (Fig. 1a). Assuming $V\ll W$, the tunnel contact
preserves the rotational symmetry of the TTQD, which is thereby
imposed on the itinerant electrons in the leads. It is useful to
treat the Hamiltonian in the special basis which respects the
$C_{3v}$ symmetry, employing an approach widely used in the theory
of Kondo effect in bulk metals \cite{CqS,CoqC}.
 The Hamiltonian $H_d+H_{lead}$
is diagonal in the basis
\begin{eqnarray}
    &&d^\dag_{A,\sigma}=
(d^\dag_{1\sigma}+d^\dag_{2\sigma}+d^\dag_{3\sigma})\label{A1}/\sqrt{3}~,\\
&&d^\dag_{E_\pm,\sigma}= (d^\dag_{1\sigma}+e^{\pm
2i\varphi}d^\dag_{2\sigma}
    +e^{\pm i\varphi}d^\dag_{3\sigma})/\sqrt{3}~; \nonumber\\
&&c^\dag_{A,{\bf{k}}\sigma}
 =
  \left(
      c^\dag_{1{\bf{k}}\sigma}+
      c^\dag_{2{\bf{k}}\sigma}+
      c^\dag_{3{\bf{k}}\sigma}
 \right)/\sqrt{3}~, \label{partw}\\
&& c^\dag_{E(\pm),{\bf{k}}\sigma}
 =
  \left(
      c^\dag_{1{\bf{k}}\sigma}+
      e^{\pm 2i\varphi}c^\dag_{2{\bf{k}}\sigma}+
      e^{\pm i\varphi}c^\dag_{3{\bf{k}}\sigma}
 \right)/\sqrt{3}~.
 \nonumber
 \end{eqnarray}
Here $\varphi=2\pi/3$, while $A$ and $E$ form bases for two
irreducible representations of the group $C_{3v}$. The Hamiltonian
of the isolated TTQD in this charge sector has six eigenstates
$|DA\rangle, |DE\rangle$. They correspond to a spin doublet ($D$)
with fully symmetric "orbital" wave function $(A)$ and a quartet
doubly degenerate both in spin and orbital quantum numbers ($E$).
The corresponding single electron energies are,
\begin{equation}\label{energy-1}
E_{DA}=\epsilon+2W, ~~~~~~~~~~~E_{DE}=\epsilon-W.
\end{equation}
 To describe the orbital effect of an external magnetic field $B$
(perpendicular to the TTQD plane and inducing a flux $\Phi$
through the triangle), one rewrites the spectrum as
\begin{equation}\label{emag}
E_{D\Gamma}(p)=\epsilon-2W\cos \left(p-\frac{\Phi}{3}\right).
\end{equation}
such that for negative $W$ and for $B=0$, $p=0,~2\pi/3,~4\pi/3$
correspond respectively to $\Gamma=A,E_\pm$.
 Fig. 2
illustrates the evolution of $E_{D\Gamma}(\Phi)$ induced by {\it
B}. Variation of $B$ between zero and $B_0$ (the value of $B$
corresponding to the quantum of magnetic flux $\Phi_0$ through the
triangle) results in  multiple crossing of the levels
$E_{D\Gamma}$.

The accidental degeneracy of spin states induced by the magnetic
phase $\Phi$ introduces new features into the Kondo effect.
 In the conventional Kondo problem, the effective low-energy exchange
Hamiltonian has the form $J {\bf S} \cdot {\bf s}$, where ${\bf
S}$ and ${\bf s}$ are the spin operators for the dot and lead
electrons respectively \cite{Andrg}. Here, however, the low-energy
states of TTQD form a multiplet characterized by both spin and
orbital quantum numbers. The effective exchange interaction
reflects the {\it dynamical symmetry} of the Hamiltonian $H_d$
\cite{KAv,KKAv}. The corresponding dynamical symmetry group is
identified not only by the operators which commute with the
Hamiltonian but also by operators inducing transitions between
different states of its multiplets. Hence, it is determined by the
set of dot energy levels which reside within a given energy
interval (its width is related to the Kondo temperature $T_{K}$).
Since the position of these levels is controlled by the magnetic
field, we arrive at a remarkable scenario: Variation of a magnetic
field determines the dynamical symmetry of the tunneling device.
Generically, the dynamical symmetry group which describes all
possible transitions within the set $\{DA,DE_\pm\}$ is $SU(6)$.
However, this symmetry is exposed at too high energy scale $\sim
W$, while only the low-energy excitations at energy scale $T_K\ll
W$ are involved in Kondo tunneling. It is seen from Fig. 2 that
the orbital degrees of freedom are mostly quenched, but the ground
state becomes doubly degenerate both in spin and orbital channels
around $\Phi=(2n+1) \pi,~~~(n=0,\pm 1,...)$.

Next we analyse the field dependent Kondo effect variable
degeneracy. It is useful to generalize the notion of localized
spin operator ${ S}^i= |\sigma\rangle \hat \tau_i \langle\sigma'|$
(employing Pauli matrices $\hat\tau_i~ (i=x,y,z)$) to
${S}^i_{\Gamma\Gamma'}= |\sigma\Gamma\rangle \hat \tau_i
\langle\sigma'\Gamma'|$, in terms of the eigenvectors (\ref{A1}).
Similar generalization applies for the spin operators of the lead
electrons: ${ s}^i_{\Gamma\Gamma'}=\sum_{{\bf k}{\bf
k'}}c^\dag_{\Gamma,{\bf{k}}\sigma}\hat \tau_i c_{\Gamma',{\bf
k'}\sigma'}$. In zero field, $\Phi=0$, the rotation degrees of
freedom are quenched at the low-energy scale. The only vector,
which is involved in Kondo cotunneling through TTQD is the spin
${\bf S}_{AA}\equiv {\bf S}$. Applying Schrieffer-Wolff (SW)
procedure, the effective exchange Hamiltonian reads,
\begin{eqnarray}\label{SWA}
 H_{SW} =
 J_E\left({\bf S}\cdot{\bf s}_{E_+E_+}+
 {\bf S}\cdot{\bf s}_{E_-E_-}\right)+
 J_{A}{\bf S}\cdot{\bf s}_{AA}
\end{eqnarray}
 The exchange vertices $J_\Gamma$ are
\begin{eqnarray}\label{coupl-Ja}
 &&J_E =-{2V^2}
   \big(\Delta^{-1}_{Q'}-\Delta^{-1}_Q\big)/{3}, \label{new-J}\\
  &&J_{A}= {2V^2}
   \big(3\Delta^{-1}_1+\Delta^{-1}_Q+2\Delta^{-1}_{Q'}
   \big)/{3},\nonumber
\end{eqnarray}
with $\Delta_1=\epsilon_F-\epsilon$, $\Delta_Q=\epsilon+Q-
\epsilon_F$, $\Delta_{Q'}=\epsilon+Q'- \epsilon_F$. Note that
$J_A>0$ as in the conventional SW transformation of the Anderson
Hamiltonian. On the other hand, $J_E<0$ due to the inequality
$Q\gg Q'$. Thus, two out of three available exchange channels in
the Hamiltonian (\ref{SWA}) are irrelevant. As a result, the
conventional Kondo regime emerges with the doublet $DA$ channel
and a Kondo temperature,
\begin{eqnarray}
T_{K}^{(A)}=D\exp\left\{-1/j_A \right\},\label{TKA}
\end{eqnarray}
where $j_A=\rho_0 J_A$, $\rho_0$ being the density of electron
states in the leads.

 At  $\Phi=(2n+1)\pi$, when the ground state of TTQD becomes spin
 and orbital doublet the symmetry of Kondo center is $SU(4)$. This kind of
{\em orbital} degeneracy is different from that of {\em
occupation} degeneracy studied in double quantum dot systems
\cite{Bord03}. The 15 generators of $SU(4)$ include four spin
vector operators ${\bf S}_{E_aE_b}$ with $a,b=\pm$ and one
pseudospin vector ${\boldsymbol {\cal T}}$ defined as
\begin{eqnarray}
{\cal T}^{+}&=&\sum_\sigma |E_+,\sigma\rangle\langle
E_-,\sigma|, \ \ {\cal T}^{-}=[{\cal T}^{+}]^{\dagger},\\
{\cal T}^z&=&\frac{1}{2}\sum_\sigma\left(
|E_+,\sigma\rangle\langle E_+,\sigma|-|E_-,\sigma\rangle\langle
E_-,\sigma| \right).\nonumber
\end{eqnarray}
Its counterpart for the lead electrons is
$
 \tau^{+}=\sum_{k\sigma}
         c^{\dag}_{E_{+}k\sigma}c_{E_-k\sigma},
         \tau_z$=$\frac{1}{2}
  \sum_{k\sigma}
  (c^{\dag}_{E_{+}k\sigma}
       c_{E_{+}k\sigma}$-$c^{\dag}_{E_-k\sigma}c_{E_-k\sigma}).$

The SW Hamiltonian is \cite{Zarand},
\begin{eqnarray}
&&H_{SW}=J_1({\bf S}_{E_+E_+}\cdot {\bf s}_{E_+E_+}+{\bf
S}_{E_-E_-}\cdot {\bf s}_{E_-E_-})\nonumber\\
&&+J_2({\bf S}_{E_+E_+}\cdot {\bf s}_{E_-E_-}+{\bf
S}_{E_-E_-}\cdot {\bf
s}_{E_+E_+})\nonumber\\
&&+J_3({\bf S}_{E_+E_+}+{\bf S}_{E_-E_-})\cdot {\bf
s}_{AA}\nonumber\\
&&+J_4({\bf S}_{E_+E_-}\cdot {\bf s}_{E_-E_+}+{\bf
S}_{E_-E_+}\cdot {\bf
s}_{E_+E_-})\label{SWE}\\
&&+J_5({\bf S}_{E_+E_-}\cdot ({\bf s}_{AE_-}+{\bf s}_{E_+A})+H.c.)
+ J_6 \boldsymbol {\cal T}\cdot{\boldsymbol \tau}
,\nonumber
\end{eqnarray}
where the coupling constants are
$J_{1}=J_{4}=J_A,~J_2=J_3=J_5=J_E$ defined in (\ref{coupl-Ja}) and
$J_6= V^2(\Delta_1^{-1}+\Delta_{Q'}^{-1})$.
Thus, spin and orbital degrees of freedom of TTQD interlace in the
exchange terms. The indirect exchange coupling constants include
both diagonal ($jj$) and non-diagonal ($jl$) terms  describing
reflection and transmission co-tunneling amplitudes. The interplay
between spin and pseudospin channels naturally affects the scaling
equations obtained within the framework of Anderson's "poor man
scaling" procedure \cite{Andrg}. The system of scaling equations
has the following form:
\begin{eqnarray}
&& {dj_1}/{dt} =
 -[j_1^2+{j_4^2}/{2}+j_4j_6+{j_5^2}/{2}],\nonumber \\
&& {dj_2}/{dt}=
 -[j_2^2+{j_4^2}/{2}-j_4j_6+{j_5^2}/{2}],
 \nonumber \\
 &&{dj_3}/{dt}=
 -[j_3^2+j_5^2],\ \
 {dj_6}/{dt}=
 -j_6^2,
 \nonumber\\
&& {dj_4}/{dt}=
 -[j_4(j_1+j_2+j_6)+j_6(j_1-j_2)],
 \nonumber\\
&& {dj_5}/{dt}=
 -{j_5}[j_1+j_2+j_3-j_6]/{2}. \label{scaling}
 \end{eqnarray}
Here $j_i=\rho_0 J_i$, and $t=\ln \rho_0 D$. Analysis of solutions
of the scaling equations (\ref{scaling})
shows that the symmetry-breaking vertices $j_{3}$ and $j_{5}$ are
irrelevant, and the vertex $j_2$, whose initial value is negative
evolves into positive domain and eventually enters the Kondo
temperature,
\begin{equation}\label{kondoe}
T_K^{(E)}=D\exp\big\{-2\big/\big[(j_A(1+\sqrt{2})+j_E+2j_\tau)\big]\big\}.
\end{equation}
 We see from (\ref{kondoe}) that both spin
and pseudospin exchange constants contribute on an equal footing.
Unlike the Kondo Hamiltonian for ${\cal N}=3$ with $J_{abcd}=J$
discussed in Ref. [\onlinecite{Affleck}], the NFL regime is not
realized for ${\cal N}=1$ with $H_{SW}$ (\ref{SWE}). The reason of
this difference is that starting with the Anderson Hamiltonian
with finite $Q, Q'$ in (\ref{H-dot}), one inevitably obtains the
anisotropic SW exchange Hamiltonian for any ${\cal N}.$ As a
result, two of three orbital channels become irrelevant. However
$T_K$ is enhanced due to inclusion of orbital degrees of freedom,
and this enhancement is magnetically tunable. It follows from
(\ref{emag}) that the crossover $SU(2)\to SU(4)\to SU(2)$ occurs
three times within the interval $0<\Phi<6\pi$ and each level
crossing results in enhancement of $T_K$ from (\ref{TKA}) to
(\ref{kondoe}) and back.\cite{foot2}
\begin{figure}[htb]
\centering
\includegraphics[width=50mm,height=50mm,angle=0]{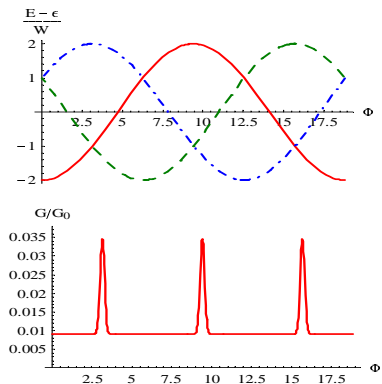}
\caption{Upper panel: Evolution of the energy levels $E_A$ (solid
line) and $E_\pm$ (dashed and dash-dotted line, resp.) Lower
panel: corresponding evolution of conductance ($G_0=\pi e^2/\hbar
$).}\label{T-phi}
\end{figure}
These field induced effects may be observed by measuring the
two-terminal conductance $G_{jl}$ through TTQD (the third contact
is assumed to be passive). Calculation by means of Keldysh
technique (at $T>T_{K}$) similar to that of Ref. \cite{kng} show
sharp maxima in $G$ as a function of magnetic field, following the
maxima of $T_K$ (lower panel of Fig. \ref{T-phi}).

\begin{figure}[htb]
\centering
\includegraphics[width=75mm,height=25mm,angle=0]{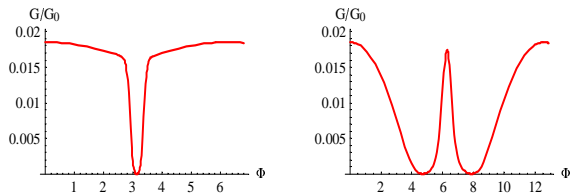}
\caption{Conductance as a function of magnetic field for
$\Phi_2=0$ (left panel) and $\Phi_1=\Phi_2=\Phi/2$ (right
panel).}\label{TQD}
\end{figure}

So far we have studied the influence of the magnetic field on the
ground-state symmetry of the TTQD. In a two-lead geometry (Fig.
1b) the field $B$ affects the lead-dot hopping phases thereby
inducing an additional AB effect \cite{footnote}. The symmetry of
the device is thereby reduced since it looses two out of three
mirror reflection axes. The orbital doublet $E$ splits into two
states, but still, the ground state is $|DA \rangle$. In a generic
situation, the total magnetic flux is the sum of two components
$\Phi=\Phi_1+ \Phi_2$. In the chosen gauge, the hopping integrals
in Eqs. (\ref{H-dot}), (\ref{H-tun}) are modified as, $W\to W\exp
(i \Phi_1/3),~~V_{1,2}\to V_s\exp[\pm i (\Phi_1/6+\Phi_2/2)],$ and
the exchange Hamiltonian now reads,
\begin{eqnarray}
H&=&J_s {\bf S}\cdot {\bf s}_s +J_d {\bf S}\cdot {\bf
s}_d
+J_{sd}{\bf S}\cdot ({\bf s}_{sd} + {\bf s}_{ds}). \label{HAB}
\end{eqnarray}
Cumbersome expressions for the exchange constants
$J_{s}(\Phi_{1},\Phi_{2})$,
 $J_{d}(\Phi_{1},\Phi_{2})$ and $J_{sd}(\Phi_{1},\Phi_{2})$ will
be presented elsewhere. They depend on the pertinent domain in
parameter space of phases $\Phi_{1,2}$. Applying poor man scaling
procedure on the Hamiltonian (\ref{HAB}) yields $T_{K}$,
\begin{eqnarray}\label{T-Kondo}
T_{K}={D}\exp\left\{-\frac{2}{j_s+j_d+\sqrt{(j_s-j_d)^2+4j_{sd}^2}}\right\},
\end{eqnarray}
and the conductance at $T>T_{K}$ reads \cite{kng},
\begin{eqnarray}
  \frac{G}{G_0}&=&\frac{3}{4}\frac{j_{sd}^2}{(j_s+j_d)^2}\frac{1}{\ln^2
  (T/T_K)}.
  \label{cond-AB}
\end{eqnarray}
The conductance $G(\Phi_{1},\Phi_{2})$ (\ref{cond-AB}) obeys the
Byers-Yang theorem (periodicity in each phase) and the Onsager
condition $G(\Phi_{1},\Phi_{2})=G(-\Phi_{1},-\Phi_{2})$. We choose
to display the conductance along two lines
$\Phi_{1}(\Phi),\Phi_{2}(\Phi)$ in parameter space of phases,
namely, $G(\Phi_{1}=\Phi,\Phi_{2}=0)$ and
$G(\Phi_{1}=\Phi/2,\Phi_{2}=\Phi/2)$ (figure \ref{TQD} left and
right panels respectively).
 The Byers-Yang relation
 implies respective periods of $2 \pi$ and $4 \pi$ in $\Phi$.
  Experimentally, the
magnetic flux is applied on the whole sample as in figure 1b, and
the ratio $\Phi_{1}/\Phi_{2}$ is determined by the specific
geometry. Strictly speaking, the conductance is not periodic in
the magnetic field unless $\Phi_{1}$ and $\Phi_{2}$ are
commensurate.

  The shapes of the conductance curves presented here are distinct from those
  pertaining to a mesoscopic AB interferometer with a single correlated
  QD and a conducting channel\cite{Hofs,AB}
  (termed as Fano-Kondo effect
\cite{Hofs}). For example, $G(\Phi)$ in Fig. 3 of Ref. \cite{Hofs}
(calculated in the strong coupling regime)  has a broad peak at
$\Phi=\pi/2$ with $G(\Phi=\pi/2)=1$. On the other hand, $G(\Phi)$
displayed in Fig. 2 (pertinent to Fig. 1a and obtained in the weak
coupling regime), is virtually flux independent except near the
points $\Phi=(2n+1) \pi$ ($n$ integer) at which the $SU(4)$
symmetry is realized and $G$ is sharply peaked. The phase
dependence is governed here by interference effects on the level
spectrum of the TTQD. The three dots share an electron in a
coherent state strongly correlated with the lead electrons, and
this coherent TTQD {\it as a whole} is a vital component of the AB
interferometer. In the setup of Fig. 1b, the Kondo cotunneling
vanishes identically on the curve $J_{sd}(\Phi_{1},\Phi_{2})=0$.
The AB oscillations arise as a result of interference between the
clockwise and anticlockwise "effective rotations" of TTQD in the
tunneling through the \{13\} and \{23\} arms of the loop (Fig.
1b), provided the dephasing in the leads does not destroy the
coherence of tunneling through the two source channels
\cite{natu}. On the other hand, in the calculations performed on
Fano-Kondo interferometers, $G(\Phi)$ remains finite \cite{Hofs}.
Another kind of Fano effect due to the renormalization of electron
spectrum in the leads induced by the lead-dot tunneling similar to
that in chemisorbed atoms \cite{Plih01} is beyond the scope of
this paper.

To conclude, we have shown that spin and orbital degrees of
freedom interlace in ring shaped quantum dots thereby establishing
the analogy with the Coqblin-Schrieffer model in real metals.
 The orbital degrees of freedom are tunable by an external
 magnetic field, and  this implies a peculiar AB effect,
 since the magnetic field affects the spectrum and the tunneling amplitudes. The
conductance is calculated in the weak coupling regime at $T>T_K$
in three- and two-terminal geometries (Figs 1a,b). In the former
case it is enhanced due to change of the dynamical symmetry caused
by field-induced level crossing (Fig. 2). In the latter case the
conductance can be completely suppressed due to destructive AB
interference in source-drain cotunneling amplitude (Fig. 3). These
results promise an interesting physics at the strong coupling
regime as well as in cases of doubly and triply occupied TTQD
\cite{KuKA}. It would also be interesting to generalize the
present theory for quadratic QD \cite{quad}, which possesses rich
energy spectrum with multiple accidental degeneracies.

\noindent
 This research is supported by grants from
 Clore foundation (T. K.), ISF (K. K., Y. A.) and DIP project (Y. A.).
 Critical comments by O. Entin-Wohlman are highly appreciated.


\begin{thebibliography}{99}
\bibitem{foot2} Real periodicity in $\Phi$ is $2\pi$: each
crossing point may be considered as a doublet $E_\pm$ by regauging
phases $\varphi$ in (\ref{partw}).
\bibitem{footnote} The necessary condition for AB effect is that
the electron coherence length in the source should exceed the size
of the electrode "tip".
\end{thebibliography}
\end{document}